# Double bosonic stimulation of THz emission in a polaritonic cascade laser

[1,2] M.A. Kaliteevski, [1] K. A. Ivanov

[1] Saint Petersburg Academic University 8/3 Khlopina Str, St Petersburg, 194021, Russia
[2] Ioffe Physicotechnical Institute of Russian Academy of Science,
26, Polytechnicheskaya,194021 St-Petersburg, Russia

**Abstract**.
We have considered a system of equidistant polaritonic states, interacting with an electromagnetic field of a localized THz cavity mode. An accumulation of photons in a THz cavity mode together with bosonic stimulation of the transition between polariton levels induces intensive radiative scattering of polaritons. The concept of double bosonic stimulation of a radiative transition in a bosonic cascade laser (BCL) is formulated, and the possibility of the use of a BCL for the emission of THz radiation is investigated. The system demonstrates threshold-like behaviour and, above threshold, cascade radiative transitions of polaritons which leads to an increase of quantum efficiency, which can exceed unity. The interaction of polaritons with a reservoir leads to an increase in the threshold pumping rate and a decrease of quantum efficiency.

**Introduction**

The creation of an efficient solid-state emitter of THz radiation remains important task of the modern physics and technology [1]. Recently, several designs of polaritonic THz emitter, based on semiconductor microcavities, were proposed, where the rate of spontaneous emission for THz photons can be additionally increased by bosonic stimulation if a radiative transition occurs into a condensate state of bosons [2-5]. Interaction of THz photons and polaritons is a subject of intensive experimental studies [6-10].

Proposed polaritonic emitters, however, have a serious disadvantage: since each polariton (which has energy of the order of 1eV) can produce only one THz photon (with energy about 10 meV) most of the pumped energy is lost and the efficiency of the device remains small. The quantum (and power) efficiency of the proposed device can be increased if the polariton is made to emit several identical THz photons. A similar effect for electrons in superlattices [11] is used in the modern quantum cascade laser [12, 13]. A system of equidistant polariton levels can be implemented if the exciton is quantized in a parabolic potential [14-15]. In the case of parabolic quantum wells, the transitions between equidistant electron levels (intersubband transition) manifests itself as peak in absorption and luminescence of TM-polarized light (pronounced even at room temperature), and the spectral position of the peak corresponds to intervals separating electron quantization levels.

Recently, the system of equidistant states of bosons was considered and a bosonic cascade laser (BCL) was proposed. The BCL demonstrates a range of intriguing properties, such as nontrivial parity-dependent effects and a quantization of the occupations of different excitonic modes [16].

The aim of this paper is to develop further the theory of BCL and investigate double bosonic stimulation of THz emission in BCL: in the presence of the cavity for THz radiation, the rate of transitions between bosonic levels goes up with an increase of both the population of bosonic states and the THz mode. Quantitative analysis is presented for the set of parameters, characterizing a realistic polaritonic microcavity interacting with THz radiation, although qualitative conclusions can be applied to a BCL of an arbitrary nature.

The scheme of the structure considered is shown in figure 1: a microcavity possessing several equidistant polaritonic levels is placed into the cavity for THz radiation [17-19] with an eigenfrequency $\omega_0$, and interval between polaritonic levels corresponding to $\omega_0$. The exciton is quantized as a whole in a specially designed potential based on a parabolic shape QW. Two possible operation regimes can be considered: strong coupling of microcavity optical mode excitons in parabolic quantum well (QW) or weak coupling. In both cases, quasiparticles (microcavity polaritons or excitons) possess integer spin and obey Bose statistics. Bose-Einstein condensation (BEC) of the polariton was recently demonstrated experimentally [20]. There is evidence that BEC of excitons can also be achieved, despite the high effective mass of excitons [21, 22]. For the both cases the probabilities of radiative transitions are in the order of $10^3$ s$^{-1}$ [2].

In the case of strong coupling of QW with $N$ exciton levels ($X_1$, $X_2$,…$X_N$) interacting with cavity mode C, the Hamiltonian matrix of the system in the basis (C, $X_1$, $X_2$,…$X_N$) reads

$$H = \begin{pmatrix} E_c & \Omega_1 & \Omega_2 & . & \Omega_N \\ \Omega_1 & E_1 & 0 & . & 0 \\ \Omega_2 & 0 & E_2 & . & 0 \\ . & . & . & . & 0 \\ \Omega_N & 0 & 0 & 0 & E_N \end{pmatrix} \quad (1)$$

where ($E_C$, $E_1$, $E_2$,…$E_N$) are the energies of the of cavity mode and exciton states; $\Omega_i$ are the values of Rabi splitting of cavity mode and exciton states with index $i$. If the equidistant excitonic states interact with cavity mode, forming N+1 polariton states with populations ($P_0$, $P_1$, $P_2$,…$P_N$), energies of polariton states, which can be obtained by diagonalization of matrix (1), will not be separated equally. Nevertheless energies of polariton levels can be made equidistant by numerical optimization [23] of exciton level energies ($E_C$, $E_1$, $E_2$, …,$E_N$) and corresponding design of the quantum well potential profile. In the case of weak coupling, a simple parabolic shape of the potential of the QW can be used, though the cavity could serve for more efficient pumping of highest exciton state in the cascade.

Dynamics of the population of equidistant polariton levels, THz photons and reservoir, when the highest polariton level is resonantly pumped with rate $I$, can be described by the system of Boltzmann equation:

$$\dot{P}_i = I\delta_{iN} - P_i/\tau_i + W_i(Q)(1-\delta_{iN})\left[P_{i+1}(P_i+1)(T+1) - P_i(P_{i+1}+1)T\right] + \\ + W_i(Q)(1-\delta_{i0})\left[P_{i-1}(P_i+1)T - P_i(P_{i-1}+1)(T+1)\right] - (1-\delta_{i0})P_i/\tau_i^- + (P_i+1)R/\tau_i^+ \quad (2a)$$

$$\dot{T} = -T/\tau(Q) + \sum_{i=0}^{N-1} W_i(Q)\left[-P_{i+1}(P_i+1)T + P_i(P_{i+1}+1)(T+1)\right] \quad (2b)$$

$$\dot{R} = \sum_{i=0}^{N}\left[P_i/\tau_i^- - (P_i+1)R/\tau_i^+\right] - R/\tau_R \quad (2c)$$

where $I$ is the pumping rate, $T$, $R$ and $P_i$ ($i$ counts from 0 to $N$) are the populations of THz mode, the reservoir and the polariton levels, $W_i(Q)$ is the probability of polaritonic radiative transition from level $i$ to level $i$-1 (see figure 1); $1/\tau_i^+$ and $1/\tau_i^-$ are the rates of phonon assisted transitions from polartion mode with index "$i$" to reservoir and back respectively; $\tau$, $\tau_R$, and $\tau_i$ is the lifetime of THz mode, and Q is quality factor of the cavity for THz radiation. The role of the THz cavity is two-fold: first, it increases the lifetime of THz photons as $\tau = \tilde{\tau}Q$, (where $\tilde{\tau}$ is the lifetime of THz photons without cavity) and second, it increases the transition probability $W_i$ due to the Purcell effect: $W_i(Q) = W_{i0}F_P$ [19]. Since the inverse lifetimes of the polariton are of a comparable frequency to the THz mode $\omega_0$, the Purcell factor reads $F_P \approx \xi\omega_0(\omega_0/Q + \tau_i^{-1} + \tau_f^{-1})^{-1}$ where $\xi$ is geometric factor and $\tau_i$ and $\tau_f$ are lifetimes of initial and final polariton states [2]. For realistic values of polariton lifetimes (picoseconds) and $\omega_0$ corresponding to THz band, $F_P$ cannot exceed $10^2$. Subsystems of polaritons and THz photons experience mutual influence: on the one hand radiative of polaritons populate the THz mode, on the other hand THz photons lead to increase of a rate of stimulated scattering of polaritons. In other words, there is "double bosonic stimulation" of polariton scattering and THz photon emission.

Figure 1 shows the dependence of occupancy numbers $T$, $R$ and $P_i$ on the pumping rate $I$ in stationary regime (when $\dot{T} = \dot{P}_i = \dot{R} = 0$) for the case of the absence of the reservoir for and $Q = 50$. The parameters used for modelling are the same as in [2]: $W_{0i} = 10^3$ s$^{-1}$, $\tilde{\tau} = 10$ ps and $\tau_i = 20$ ps, the lifetimes $\tau_i^+$ and $\tau_i^-$ describing interaction with reservoir, are chosen to be 100 ps [24].

It can be seen that the dependence of occupancy number on pumping is characterized by a threshold: below the threshold, the relaxation of polaritons is not stimulated by the THz photons and for the highest level $P_N \approx I\tau$, for the second to highest $P_{N-1} \approx WI\tau_i^2$, and for each subsequent level population is reduced by the factor $W\tau_i$. Population of the THz mode below the threshold can be estimated as $T \approx WI\tau_i\tau$. When pumping rate achieves threshold value $I_{th} \approx (W\tau_i\tau)^{-1}$, $T$ reaches unity and threshold occurs: $T$ and all $P_i$ except $P_N$ demonstrate rapid growth. An increase of T above threshold leads to stimulated radiative transitions between polariton levels, which in turn lead to a levelling of the population of different polariton states but not to an accumulation of polaritons in the ground state, as it is in the case of a usual BEC of a polariton [20].

Accumulation of photons is a crucial mechanism for the performance of a BCL, and the dependence of the threshold pumping intensity on the quality factor $Q$ is shown as an inset to Figure 2.

It can be seen that the threshold pumping rate $I_{th}$ decreases substantially with increasing Q. If the system of polaritons can interact with a condensate, as shown in figure 1, the polariton from the highest level can experience relaxation without the emission of THz photons. The latter leads to an increase of $I_{th}$, which is demonstrated by the red line in the inset in figure 2.

The dependence of the external quantum efficiency (defined as $\beta \approx T/(\tau I)$) on the pumping rate is shown in figure 3 for a series of structures with different polaritonic level numbers $N = 2,3,4,5,6$ and 7. As well as the occupation number of a photon mode $T$, quantum efficiency $\beta$ demonstrates threshold-like behaviour. Below threshold, $\beta$ has a value of the order of $W\tau_i$ which is extremely small. Note, that below the threshold, the values of $\beta$ for various $N$ are equal: since probability of THz emission $W$ is much smaller than inverse lifetime of polariton modes, each polariton emit only one THz photon regardless the number of the polariton level in the cascade. At threshold, the value of $\beta$ is increased by the factor $(W\tau_i)^{-1}$ reaching the order of unity. Above the threshold, quantum efficiency is linearly proportional to the number of the polariton level in the cascade, which is illustrated by an inset in figure 3. When the polaritonic cascade interacts in the reservoir, a non-radiative relaxation channel is opened and $\beta$ decreases, but nevertheless remains proportional to the number of levels in the cascade, as illustrated by the red line.

Finally, we should note, that the increase of quantum efficiency of a BCL is explained not only by an increased number of radiative transitions experienced by each polariton, injected to the top of cascade, but equally by stimulation of the radiative polariton transitions. To conclude, we have proposed a concept of the bosonic cascade laser. As model structure, we have considered a semiconductor structure with equidistant polariton states placed inside a cavity for THz radiation. The proposed structure can be used as an emitter of THz radiation characterized by external quantum efficiency exciding unity.

Authors are grateful to M. M. Glazov, I.A. Shelykh, A.V. Kavokin, T. Liew and A. Gallant for useful discussion. The work was supported by FP7 IRSES POLATER and POLAPHEN projects, and RFBR.

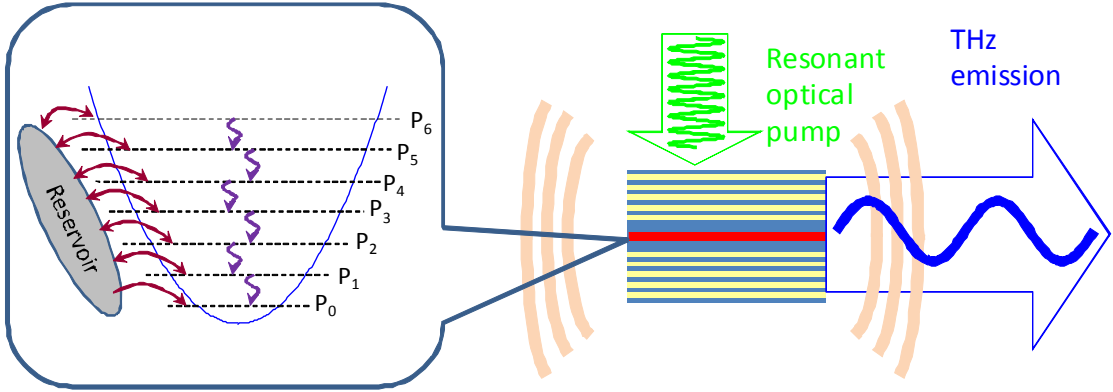

Figure 1. Right: Scheme of the structure: polaritonic microcavity is embedded into the cavity for THz radiation (shown not to scale). Left: radiative THz transitions in the system of equidistant polariton level.

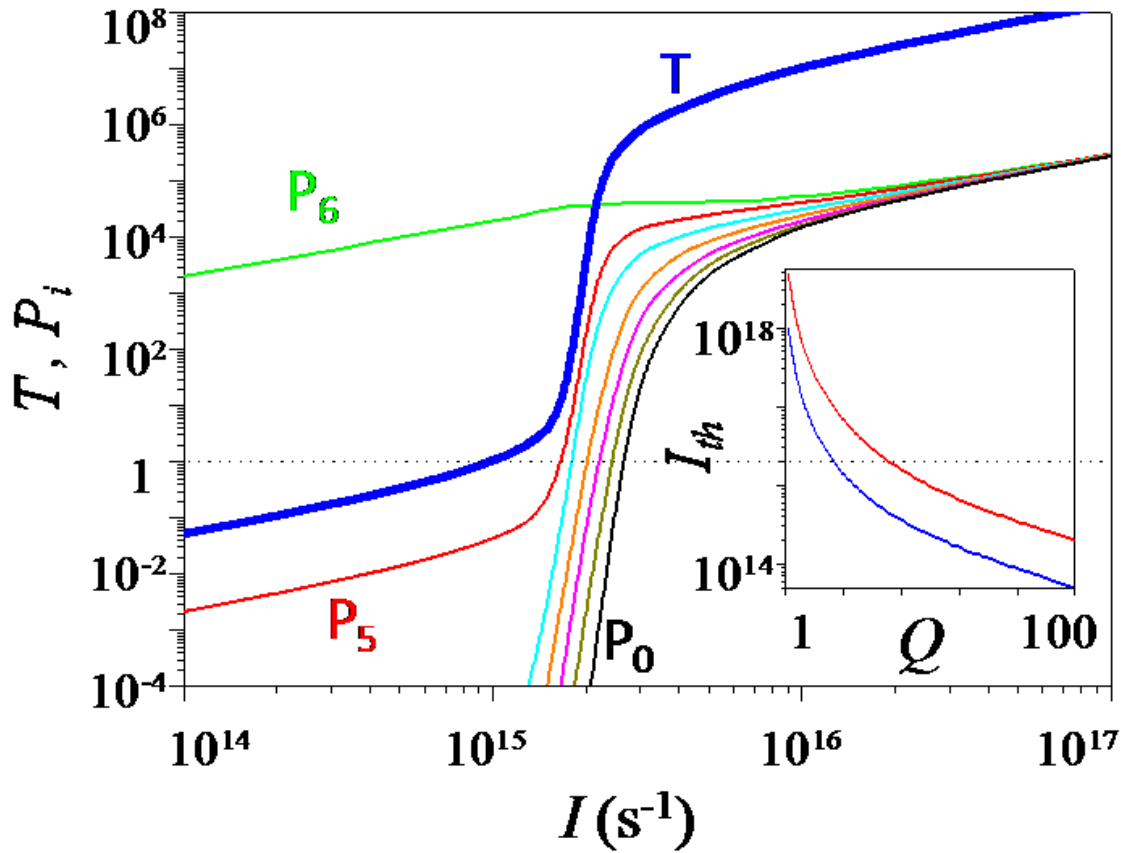

Figure 2 (colour-on-line). Populations of polaritonic levels $P_i$ and THz mode $T$ as a function of pumping in the case of absence on interaction with reservoir for the system of $N = 7$. The Inset shows threshold pumping rate as a function of quality factor of THz cavity Q without reservoir (blue line) and with reservoir (red line).

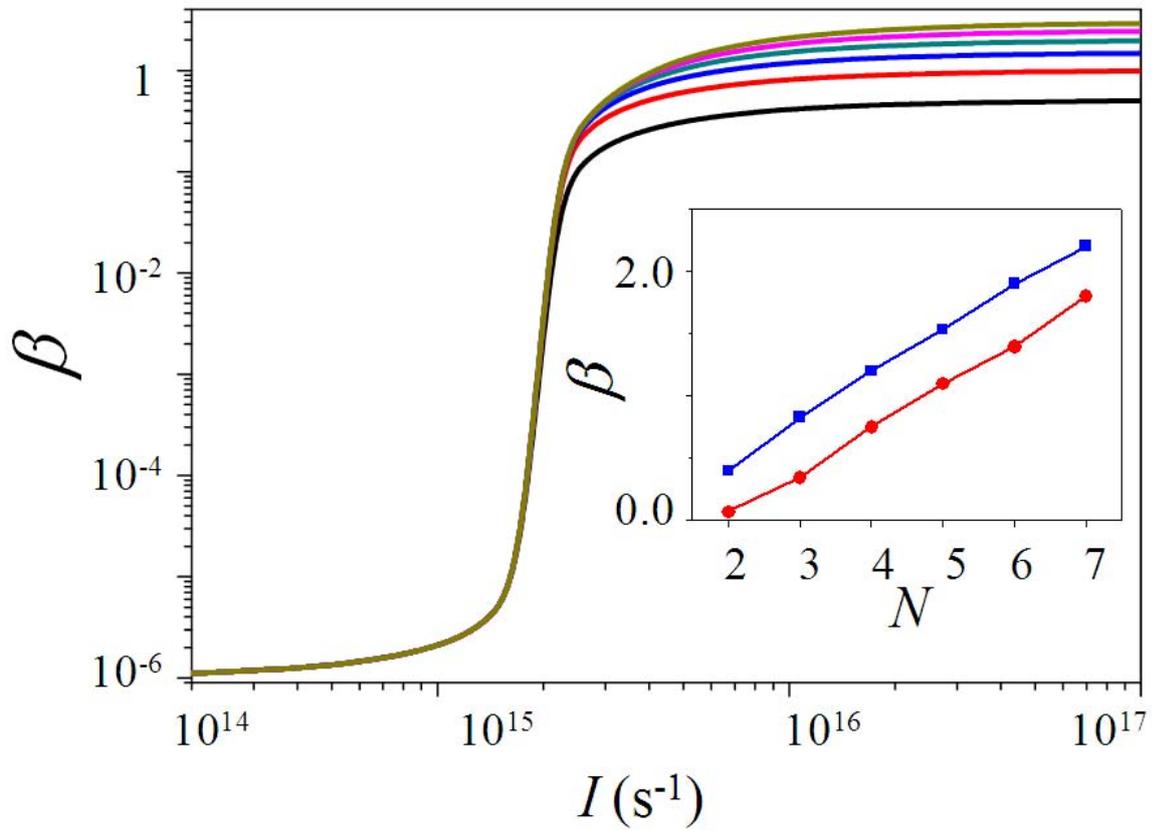

Figure 3 (colour-on-line).
External quantum efficiency $\beta$ as a function of pumping rate $I$ for different number of polaritonic levels $N$ =2, 3, 4, 5, 6, and 7. Polaritons do not interact to condensate. The inset shows a dependence of $\beta$ as a function of number of polariton level $N$ without interaction to the reservoir (blue line) and with reservoir (red line).